\documentclass[10pt,conference]{IEEEtran}
\usepackage{amsmath}
\usepackage{amsfonts}
\usepackage[utf8]{inputenc}
\usepackage{breqn}
\usepackage{enumitem}
\usepackage{wrapfig}
\usepackage{fancyvrb}
\usepackage{multirow}
\usepackage{graphicx}
\usepackage{epstopdf}
\usepackage{pdflscape}

\DeclareMathOperator*{\argmin}{arg\,min}
\DeclareMathOperator*{\argmax}{arg\,max}

\begin{document}






%

\title{Profiling Users by Modeling Web Transactions}
%
%
%
%
%

%
\author{\IEEEauthorblockN{Radek Tomšů}
       \IEEEauthorblockA{Aalto University\\
       Email: radek.tomsu@aalto.fi}  
\and     
\IEEEauthorblockN{Samuel Marchal}
       \IEEEauthorblockA{Aalto University\\
       Email: samuel.marchal@aalto.fi}
\and     
\IEEEauthorblockN{N. Asokan}
       \IEEEauthorblockA{Aalto University and University of Helsinki\\
       Email: asokan@acm.org}}

\maketitle

\begin{abstract}
Users of electronic devices, e.g., laptop, smartphone, etc. have characteristic behaviors while surfing the Web. Profiling this behavior can help identify the person using a given device. In this paper, we introduce a technique to profile users based on their web transactions. We compute several features extracted from a sequence of web transactions and use them with one-class classification techniques to profile a user. We assess the efficacy and speed of our method at differentiating 25 synthetic users on a benchmark dataset (from a major security vendor) representing 6 months of web traffic monitoring from a small enterprise network.
\end{abstract}

\section{Introduction}

Modeling network communication patterns has several applications including intrusion detection~\cite{wang2004anomalous}, bot detection~\cite{jacob2012pubcrawl} and identification of running applications on a device~\cite{conti:2016:analysing}.
While network traffic monitoring has also been used to profile host communications~\cite{hammerschmidt2016efficient,karagiannis2007profiling,Xu2005}, little attention has been given to \textit{profiling a specific user} based on the network traffic she generates. The few existing techniques~\cite{hammerschmidt2016efficient,pang2007802,verde2014no} use coarse-grained features of communication such as IP flow records. Thus, they require a long period of traffic monitoring to reliably identify the communicating user, which limits their application.

A system able to quickly identify if a known user is using a given device can have many applications such as continuous authentication or intrusion monitoring. In these scenarios, when the system detects user behavior that is inconsistent with previously seen behavior, an administrator is alerted (intrusion monitoring) or the user is automatically logged out (continuous authentication).

The patterns of behavior of people as they access various Internet sites are likely to be consistent over time for a given user while being different from one user to another.
Based on this assumption, we conjecture that a person can be identified by profiling his/her web transactions. A web transaction is a sequence of HTTP requests and responses to a single URL.
We introduce a feature extraction method applied to sequences of web transactions. We extract these features from historical logs of web transactions to build a user-specific profile using one-class classification algorithms: OC-SVM and SVDD. We use an optimization method to compute the best parameters for building these models. 
In contrast to state-of-the-art user profiling solutions~\cite{hammerschmidt2016efficient,verde2014no} relying on IP flows, we use web transaction logs augmented with information about the requested service, i.e. website category, application type, etc., to build our user profiles. Thus, our profiles are more specific enabling quick user identification within a few minutes.


\noindent\textbf{Contributions:}
We introduce \textit{\textbf{a novel feature extraction and modeling technique}} to profile users from their web transactions (Sect.~\ref{sec:methodology}).
We evaluate our profiling technique with a benchmark dataset provided by a major security vendor. The dataset consists of logs collected from 25 synthetic users over a period of 6 months (Sect. IV). The vendor uses this dataset for comparative evaluation of commercial user profiling solutions.
Our preliminary results show that our technique can \textit{\textbf{identify a user accurately (90\%) and quickly (\textless 5 minutes)}} (Sect.~\ref{sec:exp}).

\section{One-class classification}
\label{sec:background}
We use two one-class classification methods that try to separate normal transactions of a user (provided as input) from unusual ones (which are unknown) by creating a binary function. The objective of this function is to accept (most of) the transactions of the user for whom it was trained, while rejecting transactions of others.
These methods are described using the following prerequisites.

Given input: 
\begin{equation}
\textbf{x}_1,\cdots,\textbf{x}_l \in X
\end{equation}
where $l \in \mathbb{N}$ is a number of input samples and $\textbf{x}_i \in \mathbb{R}^n$. Let $\Phi$ be a feature mapping $X \rightarrow F$ where $F$ is an inner product space where the inner product can be computed by a kernel function $k$ e.g. Linear kernel, Sigmoid kernel, etc.~\cite{scholkopf2001estimating}:
\begin{equation}
\textbf{x}, \textbf{y} \in X  ,  k(\textbf{x},\textbf{y}) = \Phi(\textbf{x}) \cdot \Phi(\textbf{y}) , k(\textbf{x},\textbf{y}) = e^{-\Vert \textbf{x} - \textbf{y} \Vert^2/C}
\end{equation}
where $C$ is a predefined constant.

\subsection{One-Class Support Vector Machines (OC-SVMs)}
OC-SVMs is an unsupervised algorithm to find the best separating hyperplane that splits the data into two regions: one with high probability density of data, where the most data lies, and the other with the rest. We consider a version of the algorithm that allows setting $\nu$, an upper bound on the fraction of outliers (training points not accepted by the model) and a lower bound on the fraction of support vectors~\cite{scholkopf2001estimating}. Basic concepts of the method are briefly presented following the original notation.

To get a support of high-dimensional distribution we need to solve the following primal problem:
\begin{gather}
\argmin_{w \in F, \boldsymbol \xi \in \mathbb{R}^l,\rho \in \mathbb{R}} \frac{1}{2} \Vert w \Vert^2 + \frac{1}{\nu l} \sum_i \xi_i - \rho \nonumber \\
\text{subject to} \hspace{10pt} (w \cdot \Phi(\textbf{x}_i)) \geq \rho - \xi_i, \hspace{10pt} \xi_i \geq 0, i = 1,\cdots,l
\end{gather}
We are optimizing a hyperplane $w$ which separates the data into two spaces with a margin at least $\rho$. In case of non-linearly separable data, there is a slack variable $\xi$, which enables some of the training points to have a smaller margin than $\rho$. When $\xi_i$ is positive, the margin of the point is lower than $\rho$. The sum of slacks should be minimized, it favors thus well separated hyperplane solution.

The decision function for the primal problem is:
\begin{equation}
f(\textbf{x}_i) = sgn((w \cdot \Phi(\textbf{x}_i)) - \rho)
\end{equation}
While non-zero slack variables $\xi_i$ are penalized (outliers are penalized), we can expect that the most input data points $\boldsymbol x_i$ will have a decision function value $f(\textbf{x}_i) \geq 0$.

We do not directly search a mapping $\Phi$ in which our data would be linearly separable but consider the following dual problem using kernels:
\begin{gather}
\argmin_{\boldsymbol \alpha} \frac{1}{2} \sum_{ij} \alpha_i \alpha_j k(\textbf{x}_i, \textbf{x}_j) \nonumber \\
\text{subject to} \hspace{10pt} 0 \leq \alpha_i \leq \frac{1}{\nu l}, \hspace{10pt} \sum_i \alpha_i = 1, i = 1,\cdots,l
\end{gather}
and its decision function:
\begin{equation}
f(\textbf{x}) = sgn \left( \sum_i \alpha_i k(\textbf{x}_i, \textbf{x}) - \rho \right)
\end{equation}

All the points satisfying $\alpha_i > 0$ are called support vectors and the primal problem hyperplane can be reconstructed as a linear combination of these support vectors:
\begin{gather}
w = \sum_i \alpha_i \Phi(\textbf{x}_i)
\end{gather}

\subsection{Support Vector Data Description (SVDD)}
SVDD is similar to OC-SVMs, though it tries to split the data into two regions by encapsulating the majority of the data in a hypersphere instead of separating it with a hyperplane. A hypersphere is defined by a center $\boldsymbol a$ and a radius $R > 0$. The volume of the hypersphere is minimized by minimizing $R^2$ and the condition that the hypersphere contains most of the training data $\boldsymbol x_i$. To allow disregarding part of the training data from the hypersphere there are added slack variables $\xi_i \geq 0$ and the weight $C$~\cite{tax2004support}.

The task is to solve the following problem:
\begin{gather}
\argmin_{R \in \mathbb{R}, \boldsymbol \xi \in \mathbb{R}^l,\boldsymbol a \in F} R^2 + C \sum_{i=1}^l \xi_i \nonumber \\
\text{subject to} \hspace{5pt} \Vert \Phi(\boldsymbol x_i) - \boldsymbol a \Vert^2 \leq R^2 + \xi_i, \xi_i \geq 0
\end{gather}

The parameter $C$ controls the number of data points lying outside the hypersphere and is related to $\nu$ in OC-SVM as $C = \frac{1}{\nu l}$. 
When the problem is solved and the center $\boldsymbol a$ and radius $R$ are computed, a tested instance $\boldsymbol x$ is detected as an outlier if:
\begin{equation}
\Vert \Phi(\boldsymbol x) - \boldsymbol a \Vert^2 > R^2
\end{equation}

The dual problem using kernels is given by:
\begin{gather}
\argmax_{\boldsymbol \alpha \in \mathbb{R}^l} \sum_{i=1}^l \alpha_i k(\boldsymbol x_i, \boldsymbol x_i) - \sum_{i,j}^l \alpha_i \alpha_j k(\boldsymbol x_i, \boldsymbol x_j)  \nonumber \\
\text{subject to} \hspace{5pt} 0 \leq \alpha_i \leq C, \sum_{i=1}^l \alpha_i = 1
\end{gather}
and the squared radius $R^2$ and decision function $f(\textbf{x})$ are computed by:
\begin{equation}
R^2 = k(\boldsymbol x_{SV}, \boldsymbol x_{SV}) - 2 \sum_{i=1}^l \alpha_i k(\boldsymbol x_i, \boldsymbol x_{SV}) + \sum_{i,j}^l \alpha_i \alpha_j k(\boldsymbol x_i, \boldsymbol x_j)
\end{equation}
\begin{dmath}
f(\textbf{x}) = sgn \left(R^2 - \sum_{i,j} \alpha_i \alpha_j k(\textbf{x}_i, \textbf{x}_j) + 2 \sum_{i} \alpha_i k(\textbf{x}_i, \textbf{x}) - k(\textbf{x}, \textbf{x}) \right)
\end{dmath} 
Support vectors are input vectors $\textbf{x}_i$ satisfying $\alpha_i > 0$. Any support vector $\boldsymbol x_{SV}$ satisfying $0 < \alpha_{SV} < C$ can be used for the calculation of $R^2$.

The algorithm implementations used in this paper are $\nu$-OC-SVM and SVDD from~\cite{chang2011libsvm}.

\section{Profiling users}
\label{sec:methodology}

\subsection{Data}
\label{data}
The data used for analysis is in the form of web transaction logs created by a secure proxy logging all user web activities. This service augments the logs with additional proprietary information, e.g. website reputation, website category, application type, etc. The logs include various information including time-stamp, requested URL, protocol, HTTP action, user-id, etc. Here is a partial example of a logged web transaction:
\begin{Verbatim}[fontsize=\small]
2015-05-29 05:05:04, www.inlinegames.com,
HTTP/1.0, GET, user_9, Games, text/html,... 
\end{Verbatim}

Out of these logs we consider the following fields:
\begin{itemize}[noitemsep,nolistsep]
\itemsep0em
\item \textit{HTTP-action}: GET, POST, CONNECT or HEAD.
\item \textit{uri-scheme}: HTTP or HTTPS.
\item \textit{category}: a description of the content pointed by the target URL, e.g., Restaurants, Phishing, Messaging, etc.
\item \textit{media\_type}: a description of the target resource, e.g., video/mp4, text/plain, audio/wav, etc.
\item \textit{application\_type}: the application running on the target resource, e.g., Rhapsody, CloudFlare, Speedyshare, etc.
\item \textit{reputation}: reputation of the URL given by the logging service, i.e. Minimal/Medium/High Risk or Unverified.
\end{itemize}

\subsection{Extracting features from web transactions}
A feature representation is adopted to use web transactions logs with OC-SVMs and SVDD. \textit{Http-action}, \textit{uri-scheme}, \textit{category}, \textit{media\_type} and \textit{application\_type} fields are represented as bag-of-words features. Bag-of-words is a simple representation transforming nominal data into numerical matrix form. For each value taken by a field of the log file, e.g. \textit{http-action}, we create a binary feature to represent its presence ($1$) or absence ($0$) in a given web transaction. For example, for the \textit{http-action} field, four binary features are created namely GET, POST, CONNECT and HEAD. \textit{Media\_type} is split into two new fields: \textit{super-type} and \textit{sub-type}, each subjected to a bag-of-words feature representation e.g.:
\begin{Verbatim}[fontsize=\small]
video/mp4 -> super-type:video, sub-type:mp4
\end{Verbatim}

In addition, we create the following features. The \textit{reputation} field is allocated to two features. Not all URL have a verified reputation, the first feature is a binary feature having the value 1 for a verified reputation and 0 for unverified. The second is a numerical value mapping the risk: Minimal = $0$, Medium = $0.5$ and High = $1$. If the reputation is unverified, this second feature is set to the default value Minimal = $0$. One feature represents if the target destination of a transaction is public (0) or private (1) i.e. internal network requests. A partial example representation of a feature vector is:
\begin{Verbatim}[fontsize=\small]
CONNECT|HTTP|reputation|verified|Messaging|..
   1     1      0         1         0
\end{Verbatim}

\subsection{Feature vector composition}
\label{fvc}
We consider aggregated sequences of transactions for modeling. We seek to reveal underlying time or order dependent structures by aggregating transactions into time windows.
We use windows moving in time by a fixed shifting factor $S$ seconds and having a duration $D$ seconds such that $S \leq D$. A window represents the transactions occurring during the period $D$. All transactions belonging to a window are aggregated into one feature vector. Binary features in transactions within a window are aggregated using a logical disjunction. Numerical features are aggregated as an average over the values from windowed transactions. The windowing is user- (user-id) or host- (source IP address) specific: only transactions from one user/host are aggregated in a given feature vector. In the following example we assume a set of transactions belonging to one window:
\begin{Verbatim}[fontsize=\small]
CONNECT|HTTP|reputation|verified|Messaging|..
   1     1       0         1         0
   0     0      0.5        1         0
   0     1       0         0         0
\end{Verbatim}
They are transformed into one feature vector:
\begin{Verbatim}[fontsize=\small]
CONNECT|HTTP|reputation|verified|Messaging|..
   1     1     0.167     0.667       0
\end{Verbatim}

\subsection{Feature vector usage}
\label{sec:fv_usage}
Each user-specific feature vector resulting from the composition discussed above represents an input sample $\textbf{x}_i \in X$ for learning an OC-SVM or SVDD model specific to this user, as described in Sect.~\ref{sec:background}.
For testing purposes, the windowing can be user- or host-specific. User-specific windowing is used to assess the accuracy of a learned model and its capability to accept future transactions from the profiled user and to reject transactions from other users. Host-specific windowing is used for real applications where we need to identify the user using a given device.


\section{Dataset and learning parameters setting}
\label{sec:dataset}

\subsection{Dataset}

Our dataset was provided by a major security vendor who uses it as a benchmark dataset for evaluation of commercial user profiling solutions. This dataset represents 6 months of web transaction logs generated programmatically in a small enterprise network. It contains 9,450,474 web transactions of 36 users on 35 different devices. Each device was used by 3 users on average (number of different devices used by a single user ranged between 1 and 17). First, we filtered out users who had fewer than 1,500 transactions, as not representative enough, and kept 25 users having from 2,514 to 4,678,488 transactions each. The median transaction count per user is 38,910. Extracting the bag-of-words features from the dataset gives a feature vector containing 843 columns described in Tab.~\ref{table_bow}.

\subsection{Consistency of user transactions over time}

For profiling users, we assume that web transactions of a given user will be consistent over time and any changes are gradual. Thus, we conjecture that by modeling transactions observed over a sufficient period of time, we can determine if new transactions are likely performed by a profiled user or not. 
To validate this assumption, we estimate the \emph{temporal novelty} in the behavior of each user as follows. We choose a point in time $t$ as the \emph{epoch delimiter} to divide the transactions of a user into two sets: \emph{observed} transactions that happened before $t$ and \emph{subsequent} transactions that would take place after $t$. We then attempt to quantify the \emph{novelty} contained in \emph{subsequent} transactions with respect to the \emph{observed} transactions.

\begin{wraptable}{r}{0.23\textwidth}
  \caption{Feature vector composition}
  \label{table_bow}
  \small
\begin{center}
  \begin{tabular}{ l | r}
    Feature category & Count \\
    \hline
    \textit{http\_action} & 4 \\
	\textit{uri\_scheme} & 2 \\
	\textit{public\_address\_flag} & 1 \\
	\textit{reputation} & 1 \\
	reputation verified & 1 \\
	\textit{category} & 105 \\
	\textit{supertype} & 8 \\
	\textit{subtype} & 257 \\
	\textit{application\_type} & 464 \\
	\hline
	\textbf{Total} & \textbf{843}\\
  \end{tabular}
\end{center}
\vspace*{-0.2in}
\end{wraptable}

\begin{figure}
	\includegraphics[width=0.5\textwidth]{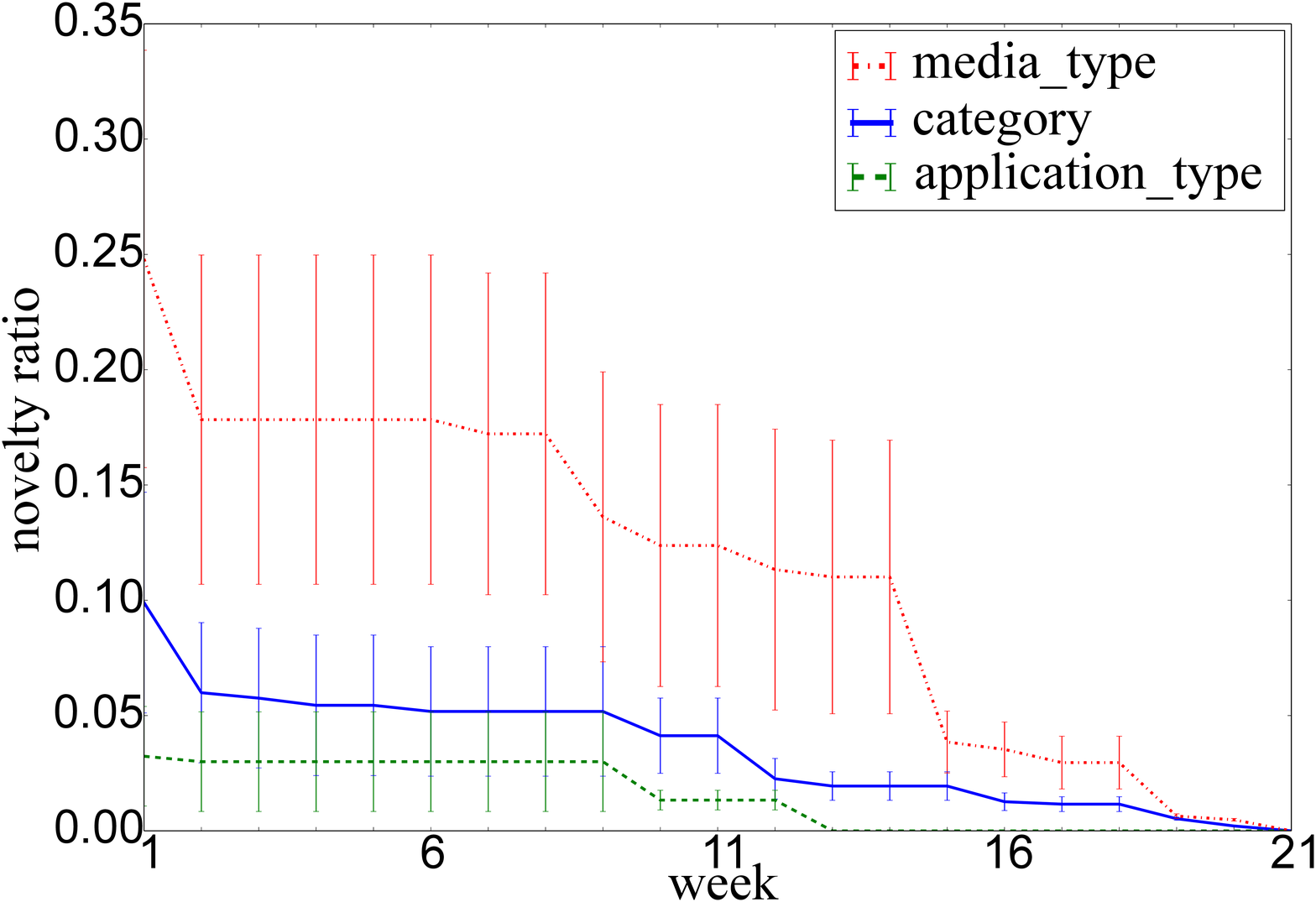}
	\caption{Novelty ratio (mean and variance) for 25 users over 20 weeks for the three largest feature categories.}
	\label{fig:consistency_big_categories}
	\vspace*{-0.1in}
\end{figure}

First, we select the three largest feature categories (cf. Tab.~\ref{table_bow}): \textit{application\_type} (464), \textit{subtype} (257) and \textit{category} (105). We compute the ratio of features from each category present in \textit{subsequent} transactions that are not in \textit{observed} transactions. This \textit{novelty ratio} represents the extent of the evolution in user behavior with respect to the dominant feature categories. Figure~\ref{fig:consistency_big_categories} depicts the evolution of this \textit{novelty ratio} for 25 users while increasing $t$ from $1$ to $21$ weeks. Already after one week of observation, there is only 25\% novel media types in the remaining of \emph{subsequent} transactions while this ratio is lower than 10\% for application types and website categories. The novelty ratio decreases as the length of the observation epoch increases to quickly reach a low value of 5\%, showing that users exhibit little novelty over time in their web transactions.
It is important to note that the reduction in novelty does indeed correspond to consistency in user behavior because the feature space covered by individual users over time is very low. The average counts of observed features per user over their whole transactions are:
\begin{itemize}
	\item \textit{category}: $17.84 / 105$
	\item \textit{subtype}: $17.12 / 257 $
	\item \textit{application\_type}: $19.08 / 464$
\end{itemize} 

\begin{figure}
	\includegraphics[width=0.5\textwidth]{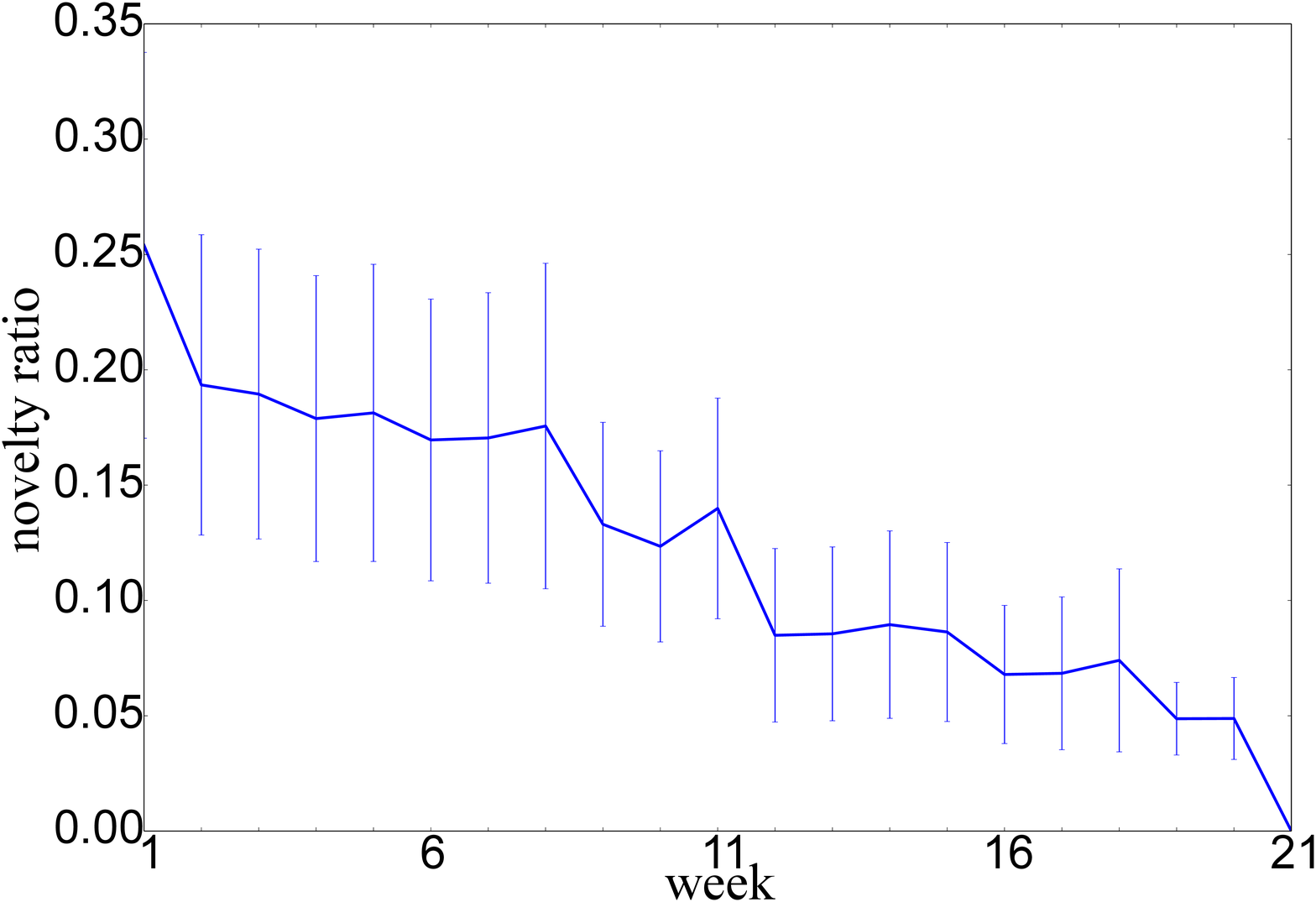}
	\caption{Novelty ratio (mean and variance) for 25 users over 20 weeks considering transaction windows.}
	\label{fig:consistency}
	\vspace*{-0.2in}
\end{figure}

Second, we evaluate the novelty in transaction windows for each user. 
We build two sets of feature vectors computed from transaction windows: one from the \emph{observed} set and one from the \textit{subsequent} set.
Then, we compute the ratio of feature vectors from the \emph{subsequent} set that are not present in the \textit{observed} set. This evaluation is more specific than the previous since feature vectors from the \textit{subsequent} set are considered as novelty if they are not strictly equal to one feature vector from the \textit{observed} set. Fig.~\ref{fig:consistency} depicts the evolution of this novelty ratio over the 25 users while increasing the observation period from $1$ to $21$ weeks. The evolution of novelty is in this case similar to that in Fig.~\ref{fig:consistency_big_categories} after a week of observations, the \emph{novelty ratio} with respect to transaction windows is 25\%.

These experiments show that user web transaction are consistent and exhibit little novelty over time. Thus, they can be modeled using our feature representation to profile users. Considering the decrease of the novelty ratio over time, we choose to use the 75\% oldest transactions from each user as \textit{training set} for parameter optimization and learning purposes. The remaining 25\% of the transactions is used as \textit{testing set}.


\subsection{Optimization of learning parameters}

To build classification models able to efficiently differentiate users, we perform a grid search to find the optimum learning parameters.
The window duration $D$, the shifting factor $S$ (Sect.~\ref{fvc}), the classification model \textit{kernel} and the parameters $\nu, C$ for OC-SVM, SVDD respectively (Sect.~\ref{sec:background}), are tested in this procedure.
We evaluate the quality of a user model along two criteria. First, the model must accept web transactions by the user the model was learned from i.e. the profiled user. The ratio of accepted transaction windows from the profiled user is called \textit{self-acceptance ratio}, denoted $ACC_{self}$, and must be maximized.
Second, the model must reject web transactions performed by others. The ratio of accepted transaction windows from other users is called \textit{other-acceptance ratio}, denoted $ACC_{other}$, and must be minimized.
The optimization problem results in maximizing the \textit{global-acceptance ratio}, $ACC = ACC_{self} - ACC_{other}$, for each user. 

Feature vectors are computed from the \textit{training set} following the user-specific windowing technique presented in Sect.~\ref{fvc}. OC-SVM and SVDD classification models are learned for each user and each combination of the learning parameters. 
The values considered for window duration $D$ and shifting factor $S$ are column headers in Tab.~\ref{table_ws_svdd}. The kernels and $\nu, C$ parameter values for model learning are column headers, row headers respectively, in Tab.~\ref{table_user_svdd}.
The window duration $D$ and shifting factor $S$ are globally optimized for all users, while the kernel and $\nu, C$ parameters are optimized per user.

Table~\ref{table_ws_svdd} provides an excerpt of the window duration and shift grid search for an SVDD model. 
$ACC_{self}$ was computed using the same transaction windows as for learning the model and $ACC_{other}$ using all other 24 users training sets.
All acceptance values $ACC_{x}$ are averages of the 25 user results. We retained a window duration $D=60s$ and a shifting factor $S=30s$ for the rest of the experiments. While these values do not provide the best global acceptance ($ACC= 79.5\%$ for $D=10m$ and $S=1m$), they provide the best self-acceptance ($ACC_{self}=93.3\%$), which is desirable to maximize user identification. In addition, keeping a low window duration and shift speeds up user identification since fewer time is needed to compute a window that would be accepted by the model. It is worth noting that the selected shifting factor $S=30s$ provides an overlap of 30 seconds between two consecutive windows and enables to compute a new feature vector every 30 seconds.

Table~\ref{table_user_svdd} presents the grid search on SVDD model parameters for $user_1$. The previously selected parameters $D=60s$ and $S=30s$ are the only considered for this part of the optimization process, which is performed individually for each user. A \textit{linear} kernel with $C=0.4$ provides the maximum \textit{global-acceptance ratio} $ACC=95.4\%$. These values are thus selected to build SVDD model for $user_1$.

\begin{table}[h]
  \caption{Excerpt of grid search results for window duration $D$ and shifting factor $S$. The results are for SVDD, $C = 0.5$ and a \textit{linear} kernel. Retained values are $D=60s$ and $S=30s$.}
  \label{table_ws_svdd}
\begin{center}
\begin{tabular}{ c || c | c | c | c | c | c}
Window duration ($D$) & 60s & \textbf{60s }& 5m & 10m & 30m & 60m\\
\hline
Shifting factor ($S$) & 6s & \textbf{30s} & 1m & 1m & 5m & 5m\\
\hline
\hline
$ACC_{self}$ & 91.1 & \textbf{93.3} & 90.1 & 90.9 & 87.6 & 83.6\\
$ACC_{other}$ & 17.2 & \textbf{15.8} & 12.7 & 11.4 & 9.6 & 8.6\\
\hline
$ACC$ & 73.8 & \textbf{77.5} & 77.3 & 79.5 & 77.9 & 75.0\\
\end{tabular}
\end{center}
\end{table}


\begin{table}[h]
  \caption{Grid search results ($ACC$) on SVDD kernel and $C$ parameter for $user_1$. $D=60s$ and $S=30s$ are fixed. A \textit{linear} kernel and $C=0.4$ are retained to build SVDD model for $user_1$.}
  \label{table_user_svdd}
\begin{center}
\begin{tabular}{ c || c | c | c | c}
$C$ $\setminus$ kernel & \textbf{Linear} & Polynomial & RBF & Sigmoid\\
\hline
0.999 & 94.4 & 0.0 & 43.6 & 85.6\\
0.99 & 94.4 & 0.0 & 43.6 & 85.6\\
0.95 & 94.4 & 0.0 & 43.6 & 85.6\\
0.9 & 94.4 & 0.0 & 43.6 & 85.6\\
0.8 & 94.4 & 0.0 & 43.6 & 85.6\\
0.7 & 94.2 & 0.0 & 42.3 & 29.4\\
0.6 & 51.6 & 0.0 & 94.1 & 85.1\\
0.5 & 94.4 & -28.6 & 43.6 & 85.6\\
\textbf{0.4} & \textbf{95.4} & 7.7 & 94.0 & 85.5\\
0.3 & 51.5 & 4.7 & 43.6 & 15.7\\
0.2 & 94.2 & -28.2 & 94.2 & 32.2\\
0.1 & 50.9 & 2.9 & 21.5 & 19.1\\
0.05 & 50.0 & 0.0 & 46.8 & 84.6\\
0.01 & 23.6 & 2.9 & 73.9 & 5.2\\
0.001 & 45.1 & 80.4 & 48.5 & 53.3\\
\end{tabular}
\end{center}
\end{table}

\begin{figure}
	\includegraphics[width=0.5\textwidth]{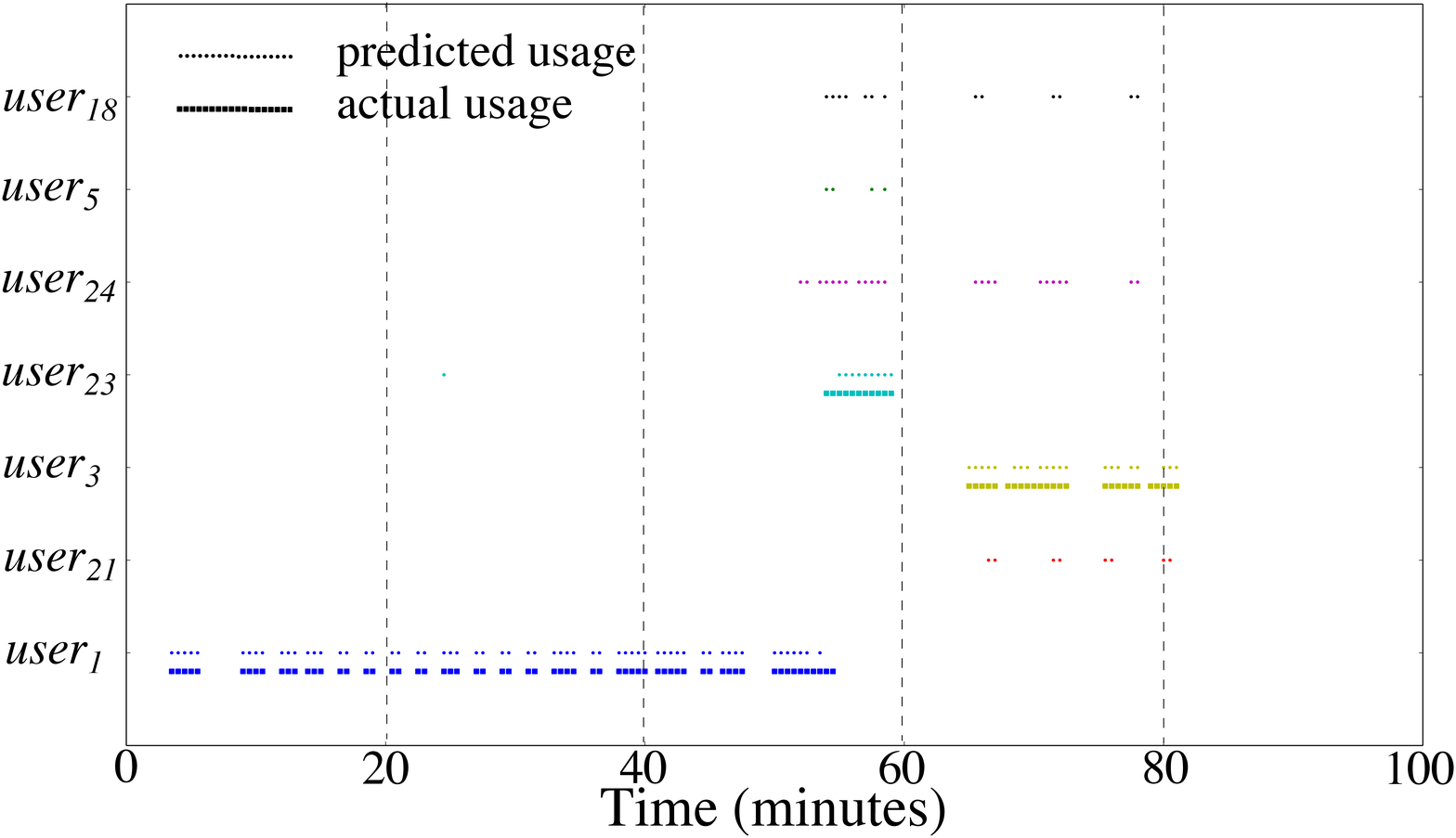}
	\caption{Transaction windows actually performed (big squared dots) by 3 users on a single device over 100 minutes. 7 out of our 25 user models accepted transaction windows (small dots) over the 100 minutes of monitoring.}
	\label{fig:identification}
\end{figure}

\section{Experiments}
\label{sec:exp}
For training, we use a window duration $D=60s$, shift $S=30s$, an individual optimized kernel and $\nu, C$ values obtained from grid search. Hence we have 25 OC-SVM and SVDD models each profiling one user.

\subsection{User differentiation}
\label{sec:user_diff}
To test the accuracy of the learned models, each user model is fed with the \textit{testing set} of all 25 users. The ratio of accepted windows per model is computed. 
Table~\ref{table_res_ocsvm} in Appendix~\ref{appendix} shows the acceptance confusion matrix for all 25 OC-SVM user models. A row represents a model $m_j$ learned from $user_j$  and applied to the \textit{testing sets} of all users. A column represents the \textit{testing set} $t_i$ from $user_i$ fed to all user models. 
A  cell gives the percentage of transaction windows from $user_i$ testing set $t_i$ accepted by the model $m_j$ from user $user_j$.

The results show generally high self-acceptance of the models ($\geq 75\%$), as observed on the diagonal, while the other-acceptance ratio is typically low. 
There are cases where a significant amount of other user transactions are accepted by a given user model, e.g. $m_{13}$ with $t_{14},t_{15},t_{16}$. This is expected since occasionally different users share similar behaviors. 
As the results for all users show in Tab.~\ref{table_avg_test}, the average self-acceptance ratio (true positive rate) of each user model is around 90\% for both OC-SVM and SVDD while the other-acceptance ratio (false positive rate) remains low: 7.3\% for OC-SVM and 10.7\% for SVDD.
This shows that the models we built are able to differentiate tens of users communicating on a same network.


\begin{table}[h]
\caption{Averaged acceptance ratio test results for OC-SVM and SVDD using individual optimized learning parameters for each user.}
\label{table_avg_test}
\begin{center}
\begin{tabular}{ c c || c | c | c | c | c | c}
\multirow{2}{*}{Window} & duration ($D$) & 60s & \textbf{60s} & 10m & 5m & 30m & 60m \\ \cline{2-8}
& shift ($S$) & 6s & \textbf{30s} & 1m & 1m & 5m & 5m \\ \hline
& $ACC_{{self}}$ & 91.7 & \textbf{89.6} & 85.9 & 87.0 & 83.7 & 81.6\\
OC-SVM & $ACC_{{other}}$ & 7.1 & \textbf{7.3} & 5.5 & 6.0 & 4.1 & 4.3\\ \cline{2-8}
& $ACC$ & 84.6 & \textbf{82.3} & 80.4 & 81.0 & 79.6 & 77.3\\ \hline
& $ACC_{{self}}$ & 91.4 & \textbf{89.4} & 92.8 & 90.7 & 85.9 & 89.7\\
SVDD & $ACC_{{other}}$ & 10.4 & \textbf{10.7} & 4.5 & 4.1 & 3.6 & 3.6\\ \cline{2-8}
& $ACC$ & 80.9 & \textbf{78.7} & 88.3 & 86.5 & 82.3 & 86.1\\
\end{tabular}
\end{center}
\end{table}

\subsection{User identification}
In this experiment we show the capability of our model to identify an unknown user on a given device. We aggregate web transactions from the testing set coming from a selected host (host-specific windowing). The obtained transaction windows are subjected to each user model to observe those that are accepted. We selected OC-SVM models for this experiment since it provided a better accuracy (lower false positives) than SVDD in Sect.~\ref{sec:user_diff}.

Figure~\ref{fig:identification} depicts the user identification process on 100 minutes of monitored web transactions from a single multi-user device. Three users are using the device over this period: first $user_1$, then $user_{23}$ and finally $user_3$ as depicted by the big squared dots in the figure, each corresponding to a transaction window. Among the 25 tested user models, seven accepted transaction windows, those are depicted by small dots in Fig.~\ref{fig:identification}.
In general, our models accurately identify the user responsible for an individual transaction window of 1 minute. Almost all transaction windows from $user_1$ are accepted only by her model. While some windows from $user_{23}$ and $user_3$ are accepted by a few other user models, we see that the longest sequences of consecutive accepted transaction windows, are for their own model.
It shows that user identification can be performed using single 1 minute transaction windows. To mitigate the effect of windows accepted by several models and increase the accuracy of the identification technique, one can consider computing a ratio of consecutive accepted windows. This would increase though the time for user identification from less than 1 minute for single windows to e.g. 5 minutes if 10 consecutive transaction windows are considered.


\subsection{Performance analysis}

\begin{figure}
	\center
	\includegraphics[width=0.45\textwidth]{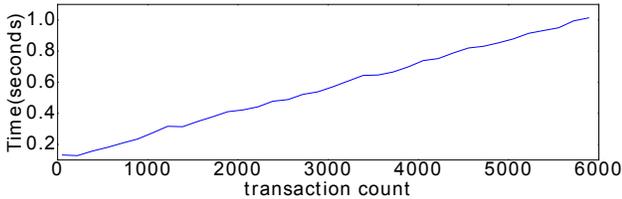}
	\caption{Speed of feature vector composition according to the number of transactions in a 1-minute window.}
	\label{fig:perf}
\end{figure}

We timed the full classification process on a desktop machine with 16GB of RAM and Intel i5-6500@3.2GHz processor. Figure~\ref{fig:perf} depicts the time taken for features extraction and feature vector composition according to the count of web transactions in the  1-minute window. The graph goes from the observed median count of transactions in one window (54) to its maximum (6,048). We see that the time consumed grows linearly with the number of transactions and that it remains below 1 second even for the biggest window we treated. Since this computation is needed every 30 seconds (sliding factor), it shows that this operation can be applied in real-time.

\begin{wrapfigure}{l}{0.2\textwidth}
	\includegraphics[width= 0.2\textwidth]{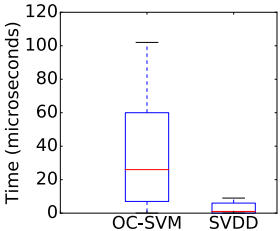}
	\caption{Prediction time.}
	\label{fig:boxwhiskers}
\vspace*{-0.2in}
\end{wrapfigure}
Once a transaction window is computed, it is classified. Figure~\ref{fig:boxwhiskers} is a box and whiskers diagram of the classification time for one window for OC-SVM and SVDD. We can see that SVDD is much faster than OC-SVM while both of them take less than $100 \mu s$ to render a decision. SVDD uses a less complex surface model representation (hypersphere) than OC-SVM to split the feature space (cf. Sect~\ref{sec:background}), which explains its faster decision. Due to the simplicity of this model, SVDD achieves slightly lower accuracy though (cf. Sect~\ref{sec:user_diff}). One would have to consider this trade-off between speed and accuracy to select either OC-SVM or SVDD for user identification.

\section{Related work}
The modeling of network communications already served several purposes including e.g. intrusion detection~\cite{wang2004anomalous}, the identification of Android applications usage~\cite{conti:2016:analysing}, IoT devices~\cite{Miettinen:2017:iot} or web crawlers~\cite{jacob2012pubcrawl}.
Most proposals focus though on modeling communications made by a single host~\cite{karagiannis2007profiling,mcdaniel2006enterprise,Xu2005} i.e. associated to a given IP address. 
In contrast, we focus in this paper on profiling users, which is more specific than host profiling, since a single host can be used by several users. Also, users exhibit more variety than applications, since a single user uses several applications and services.
To achieve user profiling, we use web transaction logs that provide more fine grained information than e.g. IP flow records used in most state-of-the-art solutions~\cite{hammerschmidt2016efficient,karagiannis2007profiling,verde2014no,Xu2005}. While few other techniques rely on full packet payload analysis for traffic modelling~\cite{wang2004anomalous,zanero2005analyzing}, we are the first to use web transaction logs augmented with service specific knowledge for profiling user.

\textsc{PubCrawl}~\cite{jacob2012pubcrawl} uses a combination of HTTP packet headers heuristics and requesting pattern analysis to distinguish robot web crawlers from human users. They build a time series counting the number of requests that arrive within consecutive 30 minutes time intervals. The web crawler detection algorithm uses, among other techniques, a \emph{Sample Auto-Correlation Function} where a strong auto-correlation at small lags indicates a regular process typical of robot crawlers. While \textsc{PubCrawl} achieves high accuracy (95\%), the binary problem of differentiating robot web crawlers from real users is a simpler task than identifying a single user out of many having more various and less predictable behaviors than robots.

Pang et al.~\cite{pang2007802} addressed the profiling of single users connected to a WiFi access point. They use 802.11 traffic characteristics such as SSID probes or broadcast packet sizes, which limit the application of their technique to WiFi connected devices.
Closer to our work is the identification of individual users behind a NAT service using IP flow data~\cite{verde2014no}.
Verde et al. extract features including the direction of a flow, the gap between two flows, the number of packets and bytes, etc.
Multiple Hidden Markov Models (HMMs) are built using these features, each representing the communications between a client and a single service e.g. \textit{www.youtube.com}. These learned HMMs are later used to see if a known user is behind a monitored NATed IP address. They achieve similar true positive rate (0.9) and false positive rate (0.08) as us. 
A drawback of this method is that it discards from analysis flows from services that were not already observed. It also requires several days of IP flow records to identify if a user is behind a NATed IP address.

Our profiling relying on URL information such as category, application and media type as features, is more general and applicable to communications with unknown services. It requires only few minutes of observation to identify if a user is using a given machine in contrast with the several hours required by most existing solutions~\cite{hammerschmidt2016efficient,jacob2012pubcrawl,pang2007802,verde2014no}.

\section{Conclusion and future work}
We introduced a feature representation that models sequences of web transactions. We showed how to use these features to profile users with OC-SVM and SVDD. We demonstrated an optimization method to compute user models and assessed these models in a user differentiation and user identification scenario using a benchmark dataset from a major security vendor. 
Our results show that our technique has 90\% true positive rate with a false positive rate of 7.3\% in transaction window classification. This method also shows promises in user identification being able to identify in few minutes which user is using a monitored device.

While the obtained accuracy and speed for single transaction windows classification are satisfying, the false positive rate we obtain needs to be improved. To increase the accuracy of our method, we will explore the inference of short-time user patterns by using only e.g. a month or a week of data for training in order to model seasonal behaviors of users. We plan to test other one-class classification algorithms e.g. auto encoders, probabilistic models. When we will reach a sufficient level of accuracy and speed, we want to develop a system for centralized continuous authentication based on web transaction monitoring. Such a solution could be implemented in a company network to assess the right usage of user accounts. We will also assess our technique on real world user data.

\section{Acknowledgments}
This work was supported in part by the Academy of Finland (grant 274951) and Intel Collaborative Research Center for Secure Computing
(ICRI-SC). We acknowledge the computational resources provided by the Aalto Science-IT project. We thank Nidhi Singh for valuable discussions.

\bibliographystyle{abbrv}
\bibliography{report_user_profiling}  

\begin{appendices}
\thispagestyle{empty}

\begin{landscape}
\section{}\label{appendix}

\vspace*{\fill}
\begin{table}[h]
  \caption{Confusion matrix for all OC-SVM user models. A table cell gives the percentage of transaction windows from $user_i$ testing set $t_i$ accepted by the model $m_j$ from $user_j$. Bold numbers in the diagonal represent the self-acceptance ratio of each user.}
  \label{table_res_ocsvm}
\begin{center}
\scalebox{0.95}{
\begin{tabular}{| c || c | c | c | c | c | c | c | c | c | c | c | c | c | c | c | c | c | c | c | c | c | c | c | c | c |}
\hline
user & $t_{1}$ & $t_{2}$ & $t_{3}$ & $t_{4}$ & $t_{5}$ & $t_{6}$ & $t_{7}$ & $t_{8}$ & $t_{9}$ & $t_{10}$ & $t_{11}$ & $t_{12}$ & $t_{13}$ & $t_{14}$ & $t_{15}$ & $t_{16}$ & $t_{17}$ & $t_{18}$ & $t_{19}$ & $t_{20}$ & $t_{21}$ & $t_{22}$ & $t_{23}$ & $t_{24}$ & $t_{25}$\\
\hline
\hline
$m_{1}$ & \textbf{93.7} & 0.0 & 0.0 & 0.0 & 0.0 & 2.0 & 0.0 & 0.0 & 0.0 & 0.0 & 0.0 & 0.0 & 0.0 & 0.0 & 0.0 & 0.0 & 0.0 & 0.0 & 0.0 & 0.0 & 0.0 & 0.0 & 2.7 & 47.8 & 0.0\\
\hline
$m_{2}$ & 0.0 & \textbf{86.8} & 22.5 & 0.0 & 0.0 & 0.0 & 0.0 & 99.1 & 0.0 & 0.0 & 0.0 & 0.0 & 0.0 & 0.0 & 0.0 & 0.0 & 0.0 & 0.0 & 0.0 & 0.0 & 0.0 & 0.0 & 0.0 & 0.0 & 0.0\\
\hline
$m_{3}$ & 0.0 & 0.0 & \textbf{75.5} & 0.0 & 0.0 & 0.0 & 0.0 & 0.0 & 0.0 & 0.0 & 0.0 & 0.0 & 0.0 & 0.0 & 0.0 & 0.0 & 0.0 & 0.0 & 0.0 & 0.0 & 0.0 & 0.0 & 0.0 & 0.0 & 0.0\\
\hline
$m_{4}$ & 0.0 & 0.0 & 0.0 & \textbf{94.9} & 0.0 & 0.0 & 0.0 & 0.0 & 0.0 & 0.0 & 0.0 & 0.0 & 0.0 & 0.0 & 0.0 & 0.0 & 0.0 & 0.0 & 0.0 & 0.0 & 0.0 & 0.0 & 0.7 & 0.9 & 0.0\\
\hline
$m_{5}$ & 0.0 & 0.0 & 0.0 & 88.1 & \textbf{97.4} & 6.0 & 0.0 & 0.0 & 0.0 & 0.0 & 0.0 & 0.0 & 0.0 & 0.0 & 0.0 & 0.0 & 0.0 & 28.6 & 0.0 & 0.0 & 0.0 & 0.0 & 0.7 & 0.9 & 0.0\\
\hline
$m_{6}$ & 0.0 & 0.0 & 0.0 & 0.0 & 5.1 & \textbf{97.5} & 0.0 & 0.0 & 0.0 & 0.0 & 0.0 & 0.0 & 0.0 & 0.0 & 0.0 & 0.0 & 0.0 & 28.6 & 0.0 & 0.0 & 0.0 & 0.0 & 20.7 & 0.0 & 0.0\\
\hline
$m_{7}$ & 0.0 & 0.0 & 0.0 & 0.0 & 0.0 & 0.0 & \textbf{93.3} & 0.0 & 0.0 & 0.0 & 0.0 & 0.0 & 0.0 & 0.0 & 0.0 & 0.0 & 0.0 & 0.0 & 0.0 & 0.0 & 0.0 & 0.0 & 0.0 & 0.0 & 0.0\\
\hline
$m_{8}$ & 0.0 & 28.3 & 0.0 & 0.0 & 0.0 & 0.0 & 0.0 & \textbf{96.6} & 0.0 & 0.0 & 0.0 & 0.0 & 0.0 & 0.0 & 0.0 & 0.0 & 0.0 & 0.0 & 0.0 & 0.0 & 0.0 & 0.0 & 0.0 & 0.0 & 0.0\\
\hline
$m_{9}$ & 0.0 & 0.0 & 0.0 & 0.0 & 0.0 & 0.0 & 0.0 & 0.0 & \textbf{100.0} & 100.0 & 0.0 & 0.0 & 0.0 & 0.0 & 0.0 & 0.0 & 0.0 & 0.0 & 0.0 & 0.0 & 0.0 & 0.0 & 0.0 & 0.0 & 0.0\\
\hline
$m_{10}$ & 0.0 & 0.0 & 0.0 & 0.0 & 0.0 & 0.0 & 0.0 & 0.0 & 43.8 & \textbf{100.0} & 0.0 & 0.0 & 0.0 & 0.0 & 0.0 & 0.0 & 0.0 & 0.0 & 0.0 & 0.0 & 0.0 & 0.0 & 0.0 & 0.0 & 0.0\\
\hline
$m_{11}$ & 0.0 & 28.6 & 0.0 & 0.0 & 0.0 & 30.8 & 0.0 & 4.3 & 0.0 & 0.0 & \textbf{100.0} & 98.6 & 0.0 & 0.0 & 0.0 & 0.0 & 0.0 & 28.6 & 0.7 & 0.0 & 45.5 & 0.0 & 26.7 & 0.0 & 0.0\\
\hline
$m_{12}$ & 0.0 & 0.0 & 0.0 & 0.0 & 0.0 & 0.0 & 0.0 & 0.0 & 0.0 & 0.0 & 0.0 & \textbf{98.6} & 0.0 & 0.0 & 0.0 & 0.0 & 0.0 & 28.6 & 2.8 & 0.0 & 0.0 & 0.0 & 35.7 & 0.9 & 0.0\\
\hline
$m_{13}$ & 0.0 & 0.0 & 0.0 & 0.0 & 0.0 & 0.0 & 0.0 & 0.0 & 0.0 & 0.0 & 0.0 & 0.0 & \textbf{96.9} & 94.8 & 87.2 & 12.2 & 100.0 & 0.0 & 0.0 & 0.0 & 0.0 & 0.0 & 0.0 & 0.0 & 10.5\\
\hline
$m_{14}$ & 0.0 & 0.0 & 0.0 & 0.0 & 0.0 & 0.0 & 0.0 & 0.0 & 0.0 & 0.0 & 0.0 & 0.0 & 96.8 & \textbf{94.2} & 74.3 & 11.3 & 100.0 & 0.0 & 0.0 & 0.0 & 0.0 & 0.0 & 0.0 & 0.0 & 10.5\\
\hline
$m_{15}$ & 0.0 & 0.0 & 0.0 & 0.0 & 0.0 & 0.0 & 0.0 & 0.0 & 0.0 & 0.0 & 0.0 & 0.0 & 4.5 & 16.8 & \textbf{83.5} & 8.1 & 100.0 & 0.0 & 0.0 & 0.0 & 0.0 & 0.0 & 0.0 & 0.0 & 9.2\\
\hline
$m_{16}$ & 0.0 & 0.0 & 0.0 & 0.0 & 0.0 & 0.0 & 0.0 & 0.0 & 0.0 & 0.0 & 0.0 & 0.0 & 71.8 & 11.6 & 16.5 & \textbf{90.0} & 0.0 & 0.0 & 0.0 & 0.0 & 0.0 & 0.0 & 12.4 & 0.0 & 71.7\\
\hline
$m_{17}$ & 0.0 & 0.0 & 0.0 & 0.0 & 0.0 & 0.0 & 0.0 & 0.0 & 0.0 & 0.0 & 0.0 & 0.0 & 69.9 & 12.1 & 60.6 & 8.7 & \textbf{100.0} & 0.0 & 0.0 & 0.0 & 0.0 & 0.0 & 9.5 & 0.0 & 0.0\\
\hline
$m_{18}$ & 0.0 & 0.0 & 17.8 & 0.0 & 0.0 & 0.0 & 0.0 & 0.0 & 0.0 & 0.0 & 0.0 & 99.5 & 0.7 & 2.2 & 3.7 & 0.3 & 0.0 & \textbf{85.7} & 8.4 & 0.8 & 0.0 & 50.0 & 51.4 & 22.8 & 0.0\\
\hline
$m_{19}$ & 0.0 & 0.3 & 0.0 & 96.6 & 2.6 & 32.8 & 0.0 & 0.0 & 0.0 & 0.0 & 100.0 & 100.0 & 2.9 & 9.4 & 14.7 & 3.3 & 0.0 & 42.9 & \textbf{88.1} & 28.0 & 45.9 & 50.0 & 2.6 & 9.1 & 1.3\\
\hline
$m_{20}$ & 0.0 & 0.0 & 0.0 & 0.0 & 0.0 & 0.0 & 0.0 & 0.0 & 0.0 & 0.0 & 0.0 & 0.0 & 0.0 & 0.0 & 0.0 & 0.0 & 0.0 & 0.0 & 0.0 & \textbf{59.2} & 0.0 & 0.0 & 0.0 & 0.0 & 0.0\\
\hline
$m_{21}$ & 0.0 & 29.7 & 20.9 & 0.0 & 0.0 & 30.8 & 0.0 & 93.1 & 0.0 & 0.0 & 100.0 & 98.6 & 0.0 & 0.0 & 0.0 & 0.0 & 0.0 & 28.6 & 3.5 & 0.0 & \textbf{90.2} & 0.0 & 35.7 & 0.9 & 0.0\\
\hline
$m_{22}$ & 0.0 & 12.1 & 0.0 & 0.0 & 0.0 & 0.0 & 0.8 & 0.0 & 0.0 & 0.0 & 0.0 & 0.0 & 0.0 & 0.0 & 0.0 & 0.0 & 0.0 & 0.0 & 7.0 & 0.0 & 0.0 & \textbf{50.0} & 0.0 & 0.0 & 0.0\\
\hline
$m_{23}$ & 0.0 & 0.0 & 15.0 & 88.1 & 0.0 & 56.7 & 0.0 & 0.0 & 0.0 & 0.0 & 0.0 & 0.0 & 0.7 & 2.2 & 3.7 & 0.3 & 0.0 & 57.1 & 10.8 & 0.0 & 0.0 & 50.0 & \textbf{88.4} & 75.0 & 0.0\\
\hline
$m_{24}$ & 0.0 & 0.0 & 18.7 & 88.1 & 0.0 & 55.2 & 0.0 & 0.0 & 0.0 & 0.0 & 0.0 & 0.0 & 0.7 & 2.2 & 3.7 & 0.3 & 0.0 & 71.4 & 9.1 & 9.6 & 0.0 & 50.0 & 68.4 & \textbf{87.1} & 0.0\\
\hline
$m_{25}$ & 0.0 & 12.1 & 0.0 & 0.0 & 0.0 & 30.8 & 0.8 & 0.0 & 0.0 & 0.0 & 100.0 & 0.0 & 80.9 & 27.8 & 98.2 & 98.4 & 100.0 & 0.0 & 1.7 & 9.6 & 45.5 & 50.0 & 0.0 & 0.0 & \textbf{92.4}\\
\hline
\end{tabular}
}
\end{center}
\end{table}
\vspace*{\fill}
\end{landscape}
\end{appendices}

\end{document}